\documentclass{elsart}

\usepackage{epsfig}

\newcommand{\dmu}{\Delta\mu}

\begin{document}
\begin{frontmatter}

\title{Correlation functions and critical behaviour 
on fluctuating geometries}
\author[kinformatyka]{P.~Bialas},
\author[kfizyka]{Z.~Burda},
\author[kopenhaga]{J.~Jurkiewicz\thanksref{fperm}}

\address[kinformatyka]{Institute of Comp. Science,
Jagellonian University,\\ ul. Nawojki 11, 30-072 Krak\'ow, Poland}
\address[kfizyka]{Institute of Physics, 
Jagellonian University\\ ul. Reymonta 4, 30-059 Krak\'{o}w, Poland}
\address[kopenhaga]{The Niels Bohr Institute,\\
Blegdamsvej 17, DK-2100 Copenhagen \O, Denmark}
\thanks[fperm]{Permanent address: Institute of Physics, 
Jagellonian University, ul. Reymonta 4, 30-059 Krak\'{o}w, Poland}

\begin{abstract}
We study the two--point correlation function in the model of branched
polymers and its relation to the critical behaviour of the model. We
show that the correlation function has a universal scaling form in the
generic phase with the only scale given by the size of the
polymer. We show that the origin of the singularity of the free
energy at the critical point is different from that in the standard
statistical models. The transition is related to the change of the
dimensionality of the system.
\end{abstract}

\end{frontmatter}

The notion of correlation length is of a paramount importance in
discrete field theoretical models.  To define the continuum limit one
requires  correlation length to be infinite.  This can  be
achieved by tuning the system to the continuous phase transition. 

Consider as an example a spin system. The correlation 
length can be defined by the behaviour of 
the connected correlation function :
\begin{eqnarray}\label{gc}
G_c(r,N)=\Big\langle\frac{1}{N}\sum_{i,j}^N \delta_{|i-j|,r}
(s_i-\langle s\rangle_N)(s_j-\langle s\rangle_N)\Big\rangle_N
\end{eqnarray}
where the averaging goes over spin configurations on the lattice of
size $N$.  The delta function selects pairs at a distance
$r$.  This function has a
well defined thermodynamic (large $N$) limit.  In this limit near
the transition point $G_c(r)$ behaves as
\begin{eqnarray}\label{gcscal}
G_c(r)\sim\frac{1}{r^{\eta-1}}u\Big(\frac{r}{\xi}\Big)
\end{eqnarray}
where $\xi=\xi(\beta)$ is the correlation length which depends
on a relevant coupling constant $\beta$ and $u(x)$ is 
a  quickly decreasing universal function, typically an exponential.
In the standard theory of continuous phase transitions the 
behaviour of $\xi$ near the critical point is characterised
by the scaling exponent~$\nu$:
\begin{eqnarray}
\xi(\beta)\sim |\Delta \beta|^{-\nu}
\end{eqnarray}
where $\Delta \beta = \beta - \beta_{c}$ is the deviation from the critical
value $\beta_{c}$. 
Integrating the connected correlation function 
(\ref{gc}) over $r$ one obtains magnetic susceptibility~:
\begin{eqnarray}
\sum_r G_c(r)=\frac{\langle M^2\rangle-\langle M\rangle^2}{N} \equiv\chi
\end{eqnarray}
where $M=\sum_i s_i$ is the total magnetization.  The susceptibility
is a second derivative of the free energy density $f(\beta,h)$ with
respect to the magnetic field $h$ and  has a singularity with a
critical exponent $\gamma$~:
\begin{eqnarray}
\chi(\beta) =\partial^2_{h} f(\beta,h) \sim |\Delta\beta|^{-\gamma}
\end{eqnarray}
Comparing with (\ref{gcscal}) one obtains
\begin{eqnarray}
\chi  \sim \xi^{-\eta+2}\sim |\Delta\beta|^{(\eta-2)\nu}
\end{eqnarray}
which leads to the Fisher scaling relation~:
\begin{eqnarray}
\gamma = (2-\eta)\nu
\end{eqnarray}
Similar reasoning can be applied to the energy--energy correlation
function, which gives specific heat $c_V$ when integrated over $r$.
In this case one obtains the scaling relation $\alpha = 2-\nu d$ 
for the critical exponent of 
$c_V = \partial^2_\beta f \sim |\Delta\beta|^{-\alpha}$.

In other words the singularity of the free energy is directly related
to the divergence of the correlation length. This picture is closely 
tied with the renormalization group analysis where infinite correlation length 
implies the scale invariance and hence the existence of a critical fixed point.
At the critical point the free energy scales 
uniformly in relevant couplings and this yields singularity of the
free energy.

In this paper we discuss a different source of the singularity 
of the free energy. We investigate a model with a random geometry. 
In such models the lattice is dynamical. 
There is an ensemble of lattices and one sums over all lattices 
in the ensemble. Geometry is defined by providing
the distance definition between nodes.

The correlation function is defined in the same 
way as in the equation (\ref{gc}).
Now the average is taken over the ensemble of lattices with size $N$, 
say with $N$ nodes.  
If there were fields on  lattices one should additionally 
average over them, but here we consider the simplest
model without the field dressing. 
One should note here that the correlation function 
in this case is not just a two-point function 
but rather a global correlator since the distance 
in the delta function depends on the whole geometry. 
Contrary to the quenched geometry models
the delta function cannot 
be pulled outside the average brackets in (\ref{gc}).

In the following we consider the branched polymer 
model \cite{adfo,bb,jk}. The
partition function in this model is given as a weighted sum over an
ensemble of trees.  Trees are weighted by one-vertex branching
weights. The partition function for the ensemble of  
trees with $N$
vertices is given by~:
\begin{eqnarray} 
Z_N=\sum_{T\in {\mathcal T}_N} \frac{1}{C(T)}
\exp\left(-\beta\sum_{i\in T} s(q_i)\right)
\label{zn}
\end{eqnarray}
where $q_i$ is order of vertex $i$ and $s$ is a one vertex action
and $C(T)$ an appropriate symmetry factor of the graph.
For some actions $s(q)$ the system has a phase transition. 
As an example we consider here the model with the action~:
\begin{eqnarray}\label{asymp} 
s(q) = \log(q)
\end{eqnarray}
The model has been solved in \cite{bb}. It has a fourth order phase
transition at $\beta_c = 2.47875... $. We discuss this particular form
of the action to fix attention but the presented results 
concerning the universality and the critical behaviour hold
also for a much broader class of actions  where 
(\ref{asymp}) is satisfied only asymptotically for large $q$ and which
in particular allow for tuning 
the critical exponents and the order of the transition with the
help of one effective parameter $\beta_c$ \cite{jk}.

The second derivative of the free energy with respect to 
$\beta$ is finite but has a singularity~:
\begin{eqnarray}
f'' \sim |\Delta\beta|^{\frac{3-\beta_c}{\beta_c-2}}
\label{f4}
\end{eqnarray}
for small $\Delta \beta$. The model has two phases~: 
the tree phase where the Hausdorff dimension is two and the
bush phase where it is infinite.  At the transition the Hausdorff
dimension is $d_H = (\beta_c-1)/(\beta_c-2)$.  The entropy exponent
$\gamma_s$ changes from the generic value $1/2$ in the tree phase to
$2-\beta_c$ in the bush phase and $\gamma_s = 1/d_H$ at the
transition \cite{jk}.

The tree phase has highly universal properties. The volume--volume correlation
function~:
\begin{eqnarray}\label{vvf}
V(r,N)=\Big\langle\frac{1}{N}\sum_{i,j}^N\delta_{|i-j|,r}\Big\rangle_N
\end{eqnarray}
is for large $N$ given by the scaling formula \cite{adj,b1}:
\begin{eqnarray}\label{vvs}
V(r,N)  = \sqrt{N} a \, v(x)
\label{v1}
\end{eqnarray}
where the universal function is
\begin{eqnarray}
v(x) = 2 x e^{-x^2}
\end{eqnarray}
and the scaling argument  
\begin{eqnarray}\label{uni}
x=\frac{a(r+d)}{\sqrt N}
\end{eqnarray}
with $d$  a {\em finite} shift which can be neglected in the large $N$ limit.
This form holds in the whole tree phase. The only dependence on the coupling 
is in the coefficient $a = a(\beta)$ which rescales the distance $r$.
The average distance between points on the branched polymers is
\begin{eqnarray}
\langle r \rangle_N = \frac{\sum_r  V(r,N) r} 
{\sum_r V(r,N)}  = \frac{\sqrt{N}}{a}
\label{v2}
\end{eqnarray}
The meaning of the formulae (\ref{v1}) and (\ref{v2}) 
is that the Hausdorff 
dimension is two as long as $a$ is finite. At the critical point 
$a$ diverges and the formula (\ref{vvs}) does not hold anymore. At this 
point  the Hausdorff dimension changes. 

The volume-volume correlation function (\ref{vvf}) is not a measure of
geometric fluctuations.  It is called correlation function only
by analogy. Its geometric meaning is the
average number of points at  a distance $r$ from a given point.  In
this way the function (\ref{vvf}) defines the dimensionality of the
system.  On a quenched geometry the 
the scaling (\ref{v1}) and (\ref{v2}) would give the canonical dimension.

To speak about geometric fluctuations and correlations one should
consider a connected correlation function of the type (\ref{gc}) for a
geometrical local quantity. In the branched polymer a good candidate 
for such a local quantity is any function which depends on a vertex order. 
In particular one can consider correlations between $q_i$ or between 
$s(q_i)$ \cite{b1,b2}. 
The latter is an analog of the energy--energy correlation function 
and integrated over $r$ gives the second derivative of the free energy 
with respect to $\beta$.

Using similar techniques as for $V(r)$ one can find the 
large $N$ limit of other correlators. We sketch the derivation 
in the appendix. In particular
for the two point connected correlator of
the vertex operator $s(q) = \log (q)$ we obtain~:
\begin{eqnarray}
G_c(r,N) &=&\Big\langle\frac{1}{N}\sum_{i,j}^N \delta_{|i-j|,r}
(s(q_i)-\langle s(q)\rangle_N)(s(q_j)- \langle s(q)\rangle_N)\Big\rangle_N
\nonumber\\
&\approx&\frac{1}{\sqrt{N}}\langle s(q)\rangle_{\infty}^2\,
\delta^2 a\,\,g(x)\qquad\qquad r>0,
\label{g0}
\end{eqnarray}
where the scaling variable $x$ is again given by (\ref{uni}).
The value at $r=0$ is a $N$ independent 
constant which we denote by $c=G_c(0,N)$. 
The universal function $g$ has the form~:
\begin{eqnarray}\label{gu}
g(x)= v''(x)=4 x \left(2 x^2-3\right)\exp\left(-x^2\right) \, .
\label{g1}
\end{eqnarray}
The coefficients $a,\delta$ and $c$ depend on $\beta$. 
Near the transition they  behave as~:
\begin{equation}
\begin{array}{rl}
a & \sim |\Delta \beta|^{-\frac{3-\beta_c}{2(\beta_c - 2)}} \\
\delta & \sim |\Delta \beta|^{\frac{3-\beta_c}{2(\beta_c-2)}} \\
\langle s(q)\rangle_{\infty} 
& \sim \mbox{const} + |\Delta\beta|^{\frac{1}{\beta-2}}\\
c & \sim \mbox{const} + |\Delta\beta|^{\frac{1}{\beta-2}}
\end{array}
\label{abc}
\end{equation}
One should note here that
similar correlation function defined in the grand--canonical ensemble
vanishes identically for $r>0$. It is essential that the averages, 
in particular $\langle s(q)\rangle_N$ are taken in the ensemble with a fixed number
of nodes $N$. The source of the long--range correlations lies in the
fact that for fixed $N$ the orders of vertices $q_i$ satisfy the global
constraint
\begin{equation}
\sum_{i=1}^N q_i = 2(N-1)
\end{equation}

The second derivative of the universal function 
$v''(x)$ in the formula (\ref{gu}) comes about as follows. One can split the 
correlation function (\ref{g0}) into three terms~:
\begin{eqnarray}\label{gsum}
G_c(r,N) & = & \Big\langle\frac{1}{N}\sum_{ij}^N \delta_{|i-j|,r}
            (s(q_i) s(q_j)\Big\rangle_N \nonumber\\
         && - \,  2 \langle s(q)\rangle_N 
\Big\langle\frac{1}{N}\sum_{i,j}^N \delta_{|i-j|,r} s(q_i)\Big\rangle_N \\ 
         && + \, \langle s(q)\rangle_N^2 
\Big\langle\frac{1}{N}\sum_{i,j}^N \delta_{|i-j|,r} \Big\rangle_N\nonumber
\end{eqnarray}
Each of them is calculated separately. 
For $r>0$ and large $N$ they correspond to three terms in the following sum~:
\begin{eqnarray}
G_c(r,N) &\approx& \langle s(q) \rangle_{\infty}^2 \Big\{
V\big(r + 2\frac{\delta}{a}\big) - 2 V\big(r + \frac{\delta}{a}\big) +
V\big( r \big) \Big\} \nonumber\\
&=& \langle s(q) \rangle_{\infty}^2 \sqrt{N}a
\Big\{v\Big(x+2\frac{\delta}{\sqrt{N}}\Big)-
2 v\Big(x+\frac{\delta}{\sqrt{N}}\Big)+v(x)\Big\}\\
&\approx&  \frac{1}{\sqrt{N}}\langle s(q) \rangle_{\infty}^2 a\delta^2 
v''(x) \nonumber
\label{glead}
\end{eqnarray}
where we used the fact that
\begin{equation}\label{Euler}
\langle s(q) \rangle_N = \langle s(q) \rangle_{\infty}+ O({1\over N})
\end{equation}
The interesting point which we would like to emphasize 
is that the dressing of the geodesic line by the operators 
$s(q)$ at the ends results only in the effective shift of  
the argument $r$. This {\em finite} shift $\delta$  
 gives a subleading contribution
to each of the correlators in the sum (\ref{gsum}).
However as the leading terms   cancel out only those subleading terms 
contribute
to the end result. As we show below, the singularity
of the free energy is directly related to the singularity of $\delta$. 
\begin{figure}[t]
\begin{center}
\epsfig{file=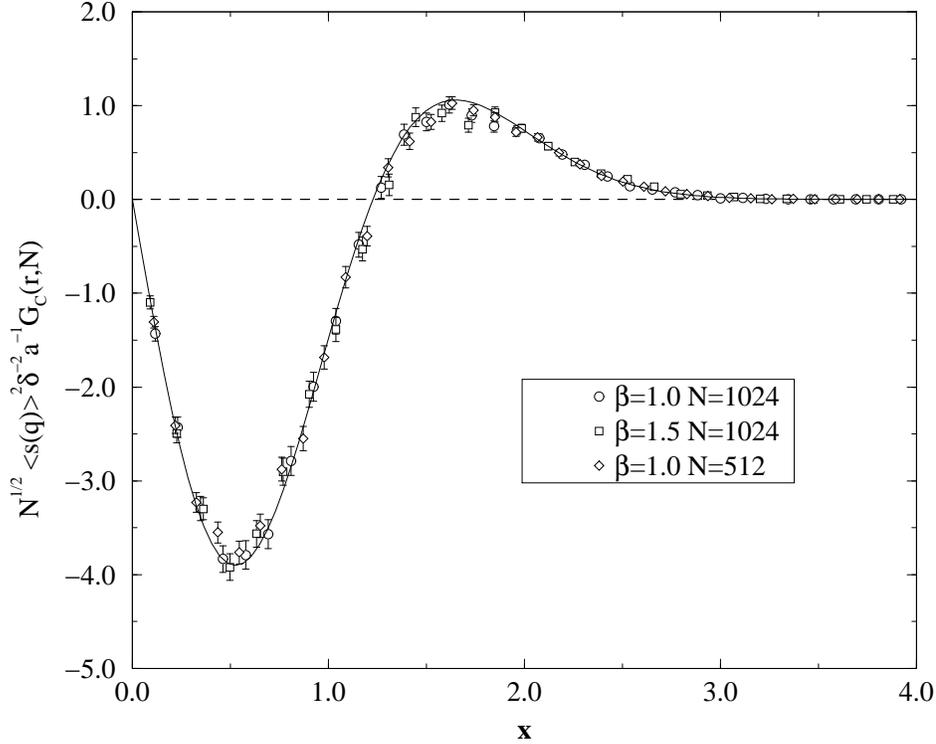,width=14cm,bbllx=12,bblly=50,bburx=576,bbury=450,
clip=true}
\end{center}
\caption{\label{fig1}Scaling behaviour of the connected correlation function.
The points are  results of 
the direct computation of $G_c(r,N)$ by the Monte--Carlo program. 
For clarity only every third point is plotted. 
The line is the universal function $g(x)$. 
The argument $x=a(\beta)r/N^{1/2}$.}
\end{figure}

An important issue is the large $N$ limit.  The formula 
(\ref{g0}) has been calculated for large $N$.
In the scaling variable $x$ (\ref{uni}) the correlation function contains
a part proportional to the delta function and the scaling part. The
scaling part gives a non--vanishing contribution when the sum
over $r>0$ is performed for $N \to \infty$ even though the contributions
for individual $r$ vanish~:
\begin{eqnarray}
\sum_{r=1}^\infty G_c(r,N) \approx \int_0^\infty dr G_c(r,N) \approx 
\langle s(q) \rangle_\infty^2 \delta^2 \int\mbox{d}x v''(x). 
\end{eqnarray}
The second derivative of
the free energy density becomes~:
\begin{eqnarray}
c_V = \partial^2_\beta f = \sum_{r=0}^\infty G_c(r,N) = c - 
2 \delta^2 \langle s(q) \rangle_{\infty}^2. 
\label{chi1}
\end{eqnarray}
The first term $c$ comes from $N$ vertices and is a trivial self
correlation.  The second one is an anomalous term coming from
correlations of pairs.  The individual correlations 
vanish like $1/N$. The number
of pairs, however, grows as $N^2$ so effectively the correlations
contribute in the same order as $c$. This of course is possible only
because the correlations are long range {\it ie.} they are not cut off
by any finite scale.  In fact it is the latter term which gives the
singularity of the free energy density. From (\ref{abc}) we see that
$a$ goes to infinity and $\delta$ to zero at the critical point.  The
singularity of $c$ is weaker than the one in $\delta^2$.  Thus the
singularity of the heat capacity $c_V$ comes from the coefficient
$\delta^2$. Paradoxically the contribution from $\delta$ vanishes at the
transition. The singularity arises from the way the coefficient
$\delta$ vanishes and leads to the divergence of higher derivatives.

In the vicinity of the critical point 
the universal argument $x = a r / N^{1/2}$ 
has two large factors~: $N^{1/2}$ and $a$ that diverge.
One has to carefully define the limit $\Delta \beta \rightarrow 0$. 
If one did it naively and sent $\Delta \beta$ to zero
at fixed $N$, the average distance $\langle r \rangle$ (\ref{v2}) 
would collapse to zero. In fact it does not.
At the transition the Hausdorff dimension changes to 
$d_H = (\beta_c-1)/(\beta_c-2)$ \cite{jk} and 
the universal argument of the functions $v$ and $g$ 
changes to $x \propto r /N^{1/d_H}$. Also the function $v$ and $g$
themselves change the form. 
Just beyond the transition point the geometry collapses. The 
collapse results from the appearance of an additional 
mass term $e^{-m(\beta) r}$ in the correlation function 
with the non--zero mass independent on $N$. This mass fixes 
the average distance between points on the polymer 
to $1/m$ which is independent of $N$. The fact that
the average distance (\ref{v2}) does not grow with the 
lattice size $N$ can be interpreted as an infinite 
Hausdorff dimension.

The model can be easily generalized to other actions. 
By a simple modification  one has a possibility to tune $\beta_c$
in the range $(2,\infty)$\cite{bb}. 
For $2<\beta_c<3$ the critical behaviour of 
the coefficients $a,\delta$ and $c$ is still described by
the same formulae (\ref{abc}). 
The coefficient $a$ diverges and $\delta$ gives the 
the singularity of the free energy. The situation 
changes for $\beta_c>3$ where both $a$, $\delta$ approach
finite constants at the critical point.
The singularities of both of them are of the form 
$\Delta \beta^{\beta_c-3}$. The singularity of $b$ 
is inherited by $c_V \sim \Delta \beta^{\beta_c-3}$.
The fact that $a$ is finite at the transition means 
that the argument $x$ of the universal functions  
(\ref{v1}) and (\ref{g1}) preserves the form 
$x = a r/N^{1/2}$ and the Hausdorff dimension is two
as in the tree phase. In  the limiting case $\beta_c=3$
the system undergoes a strong third order phase transition
with the 
singularity\footnote{not $f''\sim \log |\Delta\beta|$
as written in \cite{bb}} $f''\sim \log |\Delta\beta|^{-1}$.

Let us at the end briefly compare the critical behaviour
of the model with the standard critical phenomena of
statistical models. The general scaling Ansatz for a two-point
correlation function involves two dimensionless ratios~: $r/\xi$
and $r/N^{1/d_H}$.  
The singularity of the free energy arises when the correlation 
length $\xi$ diverges. In our model the scaling function 
involves only the  argument $r/N^{1/d_H}$. 
In other worlds it has no other scale except the size of the system.
One can say that the model is always critical ($\xi \to \infty$). 
The long--range correlations 
between orders of different vertices
are not induced by local interactions
but rather by the Euler relation (\ref{Euler}).
The mechanism of the transition is the same as in the
balls-in-boxes model and relies on appearance 
of the surplus anomaly \cite{bbj}. 
There the  correlations between number of balls in two boxes
vanish  for large $N$ (this is in the analogy with the vanishing of $G_c(r,N)$
for fixed $r > 0$ and $N \to \infty$), 
but this is compensated  by the growing number
of pairs of boxes over which we have to sum over. 
The anomaly corresponds to
the singular vertex which changes the geometrical properties
of the branched polymer and leads to the collapse which
is associated with the appearance of the non scaling mass
in the two point correlation function.

A similar phase transition between the branched polymer phase
and the collapsed phase is observed in 4d simplicial gravity
\cite{bbpt,bb2,ckr}. One can expect the same qualitative
behaviour of the correlation function (\ref{gc}) 
for the curvature-curvature correlations in 
4d simplicial gravity \cite{bs}.

Two of the  authors (P.B. \& Z.B.) thank the Niels Bohr Institute
for the hospitality during their stay,
where the work was completed. The authors are grateful to J. Ambj\o rn for
valuable comments and discussions.
The work was partially financed
by the KBN grants 2P03B04412 and  2P03B19609.

\appendix

\section{}

The correlation functions are constructed by means 
of the partition function 
of {\em planted, rooted, planar} trees. This partition
function can be found from the following recursive
equation~\cite{adfo}: 
\begin{equation}\label{zgen}
Z=e^{-\mu}F(Z)
\end{equation}

where the generating function is
\begin{equation}
F(Z)=\sum_{q=1}e^{-\beta s(q)} Z^{q-1} \, .
\end{equation}
For $\mu$ approaching a critical value $\mu_C$ 
from above, $Z$ has the following singularity~:
\begin{eqnarray}\label{zexp}
Z(\mu)=Z_0(\beta)-Z_1(\beta)\sqrt{\dmu} + \dots
\end{eqnarray}

The critical value $\mu_C$ corresponds in the large $N$ limit to
the free--energy density of the canonical ensemble
\cite{bb}. In particular
\begin{eqnarray}
\langle s(q)\rangle_N=\partial_\beta f(\beta)
= - \partial_\beta \mu_C(\beta)+O(\frac{1}{N})
\end{eqnarray}
As a function of $\beta$ the free energy $\mu_C$ 
has a singularity at a critical point $\beta_C$. This 
singularity appears also in the coefficients $Z_i$ of the
expansion (\ref{zexp})~\cite{bb,b1}~:
\begin{eqnarray}\label{sing}
1-Z_0&\sim& |\Delta\beta|^{\frac{1}{\beta_C-2}}\\ 
Z_1 &\sim & |\Delta\beta|^{\frac{3-\beta_C}{2}} 
\end{eqnarray}

The correlation functions in the grand--canonical ensemble
can be calculated in terms of $Z(\mu,\beta)$ \cite{adj,b1} as~:
\begin{eqnarray}
\sum_{T\in\mathcal T}\sum_{i,j}\delta_{|i-j|,r}&=&
\Big(1-\frac{Z}{\partial_\mu Z}\Big)^{r-1}Z^2\\
\sum_{T\in\mathcal T}\sum_{i,j}s(q_i)\delta_{|i-j|,r}&=&
\Big(1-\frac{Z}{\partial_\mu Z}\Big)^{r-1}
Z^2\frac{\partial_\beta Z}{\partial_\mu Z}\\
\sum_{T\in\mathcal T}\sum_{i,j}s(q_i)s(q_j)\delta_{|i-j|,r}&=&
\Big(1-\frac{Z}{\partial_\mu Z}\Big)^{r-1}
\Big(\frac{\partial_\beta Z}{\partial_\mu Z}\Big)^2
\end{eqnarray}

which then can be transformed by Laplace transform 
to the canonical ensemble with a fixed size $N$ leading
to the terms in (\ref{gsum}).
The results (\ref{glead}) correspond to 
the leading order terms of the Laplace 
transform calculated by the 
saddle point approximation.

\end{document}